\documentclass[twocolumn,showpacs,preprintnumbers,superscriptaddress,amsmath,amssymb]{revtex4}
\usepackage{amsfonts}
\usepackage{amsmath}
\usepackage{amssymb}
\usepackage{graphicx}%
\usepackage{footmisc}
\usepackage{bm}
\usepackage{braket}
\begin{document}
\title{Gravitational Baryogenesis Of Vacuum Inflation}
\author{Zhiqiang Huang}
\affiliation{State Key Laboratory of Magnetic Resonances and Atomic and Molecular Physics, Wuhan Institute of Physics and Mathematics,
Chinese Academy of Sciences, Wuhan 430071, China}
\affiliation{ University of the Chinese Academy of Sciences, Beijing 100049, China}

\author{Qing-yu Cai}
\thanks{Corresponding author. Electronic address: qycai@wipm.ac.cn}
\affiliation{State Key Laboratory of Magnetic Resonances and Atomic and Molecular Physics, Wuhan Institute of Physics and Mathematics,
Chinese Academy of Sciences, Wuhan 430071, China}
\date{\today}

\begin{abstract}
We show that in the vacuum inflation model, the gravitational baryogenesis mechanism will produce the baryon asymmetry. We analyze the evolution of entropy and baryon number in the vacuum inflation model. The comparison between dilution speed and the chemical potential may give a natural interpretation for decouple temperature of  the gravitational baryogenesis interaction. From the result, the mechanism can give acceptable baryon-to-entropy ratio in the vacuum inflation model.
\end{abstract}


\maketitle

\section{INTRODUCTION}\label{s1}
A model consider that the universe can be spontaneously inflated has been proposed~\cite{PhysRevD.89.083510}. Instead of inflaton, the Universe are derived by the quantum potential in this model. This model can also give the scalar and tensor perturbations after the fluctuations revision are introduced~\cite{0264-9381-34-10-105013}. Instead of inflaton field, the radiation and matter are created by Hawking radiation in this model. The reheating process is also replaced by the rapid decay of quantum potential. Though this model gives some good results. It is still a immature theory. One of the problem is the baryon asymmetry. The Hawking radiation itself can not give the the matter-antimatter asymmetry. Hence we need propose a mechanism to generate the matter-antimatter asymmetry in this model.

From the observations of cosmic microwave background radiation (CMBR) or the predications of Big Bang nucleosynthesis. The Universe contains an excess of matter over antimatter. The matter-antimatter asymmetry is the baryon or lepton asymmetry indeed. The ratio of baryons to photons from the Cosmic Microwave Background Radiation by the PLANCK experiment is~\cite{refId0} 
\begin{equation}
	\frac{n_B}{s}=0.864_{-0.015}^{+0.016}\times{10^{-10}},
\end{equation}
where $n_B$ is the density of the baryon and $s$ is the entropy density. All of the baryogenesis theory that can give acceptable baryon-to-entropy ratio are beyond the standard model. So how the baryon asymmetry was generated is remains a mystery.

It is generally accepted that the baryon asymmetry should be generated dynamically as the Universe expands and cools.  A. Sakharov listed three ingredients to realize the baryon asymmetry~\cite{Sakharov1967Violation}. (1) Baryon-number nonconserving interaction. The universe is supposed to start with zero baryon number. But the contemporary universe has nonzero baryon number. The baryon-number should not be a conserved quantity. (2) CP violation. The initial matter-antimatter symmetry means one particle always accompany with an antiparticle. If interaction does not break the charge conjugation and parity. The antiparticle will always produce the opposite particles of what particle produces at the same rate.
(3) An arrow of time (Departure from thermal equilibrium). Thermal equilibrium can recovery T symmetry. The CPT theorem tell us that it will recovery CP symmetry also. So, an arrow of time is necessary for baryogenesis.

Though the standard model can realize baryogenesis by sphaelerons~\cite{Dine2003Origin}. The CP-violating phases is too small to generate acceptable baryon number. It is necessary to seek for more effective sources of CP-violating process. One of the intriguing thought is that the gravitational interaction should be able to break any global symmetry~\cite{Peccei1998Discrete,PhysRevD.90.083535}. The baryon symmetry belongs to global symmetry. It is natural to break such symmetry in the strong gravitational system. Gravitational anomaly from index theorem is one kinds of gravitational baryogenesis model~\cite{Christensen1979New}. The Chern-Simmons term of gravity will introduce lepton number current. The couple between inflaton and Chern-Simmons term will arise the chemical potential for lepton. The chemical potential can be expressed as the change rate of inflaton. The change of inflaton during the inflation will give the chemical potential to the lepton, and therefore leptogenesis. Unfortunately, the baryon number will be suppressed by Hubble parameter. When we take the Hubble parameter of early stage of inflation. The baryon number is too small. If the sphaelerons are active after inlflation. The introduced Chern-Simmons term will change the mode of gravity wave even when they are out of the horizon. They may not be able give acceptable baryon number.

Another gravitational baryogenesis mechanism was proposed about a decade ago~\cite{Davoudiasl2004Gravitational}. It gives CP-violating interaction by the combination of the derivative of the Ricci scalar curvature $\mathcal{R}=12H^2+6\dot{H}$ and the B-current $J^\mu$
\begin{equation}\label{rb}
	\frac{1}{M_*{}^2}\int d^4\text{x}\sqrt{-g}(\partial _{\mu }\mathcal{R})J^{\mu },
\end{equation}
where $M_*$ is the cutoff scale of the effective theory. In this theory, the change rate of Ricci scalar curvature gives the chemical potential of baryon $\mu_B \sim \pm\dot{\mathcal{R}}/M_* ^2$. The highly homogeneous of the Universe can also explain the homogeneous of baryon asymmetry. During Radiation-dominated era, the Ricci scalar curvature equal to zero and not be able to give baryon asymmetry. During Matter-dominated era, the Hubble parameter is too small to give any observable phenomenon of baryon asymmetry. These means the baryon asymmetry are mainly produced during the inflation period. At this period, the slow change of Hubble parameter can gives stable production of baryon asymmetry. This means that the baryon density will not be diluted even during the inflation. With that, we can break the common sense that the baryon asymmetry process should be occurred after the inflation~\cite{Dine2003Origin}. We will adopt this mechanism and show how to realize baryogenesis in the vacuum inflation.

For simplicity, we use Planck units in this paper.

\section{Gravitational Baryogenesis Of Vacuum Inflation}
The temperature is cool down during the inflation, the Universe is obviously departure from thermal equilibrium. With CP-violating interaction Eq.~{\ref{rb}, CP violation and  Baryon-number nonconserving interaction are fitted. So all the ingredients gived by Sakharov are satisfied. In order to calculate the baryon-to-entropy ratio, we need know evolution of the baryon density and entropy density in the vacuum inflation. In the vacuum inflation model, the radiation and matter are created by Hawking radiation~\cite{0264-9381-34-10-105013}. We can calculate the entropy of them. For $\mathcal{R}$ and $\dot{\mathcal{R}}$ are nonzero during the inflation, the interaction in Eq.~(\ref{rb}) will give a chemical potential to the baryon number. It will derive the Universe towards nonzero equilibrium baryon asymmetry. During the inflation, the temperature will drop down. Once it drops below a decouple temperature $T_D$, the baryon number will be frozen and no longer change. Unlike that, the entropy will continue to increase until the end of the inflation.

At first, we will analyze the entropy, temperature and matter density evolution in the vacuum inflation model. For the temperature is very high during the inflation, all the particles created at this period are relativistic. The evolution of matter density $\rho$ is~\cite{0264-9381-34-10-105013}
\begin{equation}\label{rho}
	\dot{\rho}=g_*\frac{3\sigma H^5}{32\pi^4}-4H\rho,
\end{equation}
where $g_*$ counts the total degrees of freedom for particles, $\sigma$ is the Stefan–Boltzmann constant and $H$ is the Hubble parameter. During the inflation, the radiation is the main component. We can obtain the temperature and entropy density of the matter~\cite{PhysRevD.74.087504} with thermodynamic relations  $T=(\frac{\rho}{2\sigma g_*})^{1/4},s=\frac{8\sigma}{3}g_*T^3$. Once we know the matter density of the Universe, the temperature and entropy are very easy to calculate.

At the early stage of inflation, the quantum potential derive the Universe inflated as
\begin{equation}\label{h}
	H=\frac{l^2}{2}\frac{C1}{C2}a^{-2-p},
\end{equation}
where $p$ and $C1/C2$ is a parameter of the vacuum inflation model, $l^2=8\pi/3$. Substituting Eq.~(\ref{h}) into Eq.~(\ref{rho}), we find the matter density will increase very fast from zero to
\begin{equation}\label{rho}
	\rho=\frac{3\sigma g_* H^4}{128(-1-p)\pi^4}.
\end{equation}
With matter density we can easily know the temperature of the matter is $T=\left(\frac{3}{-1-p}\right)^{1/4}\frac{H}{4\pi}$
, which is a little different from the temperature of Hawking radiation  $T=H/(2\pi)$. The entropy density at the early stage of inflation is
\begin{equation}
	s=\sigma g_* \left(\frac{1}{12(-1-p)^3}\right)^{1/4} \frac{H^3}{\pi^3}.
\end{equation}
At the late stage of inflation, the Universe is about to exit the inflation. The change of hubble parameter is very complicate. We need use numerical calculation. The evolution of temperature of the vacuum inflation model is showed in Fig.~(\ref{entropy}).
\begin{figure}[ht]
\centering
\includegraphics[width=0.45\textwidth]{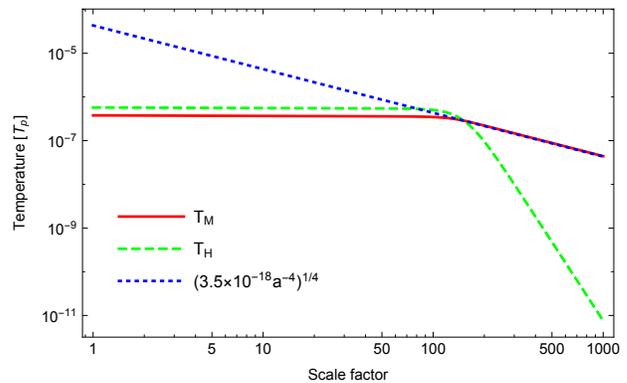}
\caption{The evolution of Hawking radiation temperature $T_H$ and the temperature of matter $T_M$. They are decrease slowly during the early stage of inflation. The quantum potential tends to zero at the end of the inflation. Hence the temperature of matter will decrease in the same form as the radiation-dominated era.} \label{entropy}
\end{figure}

Then we will calculate the baryon number. As we will see, the chemical potential will decrease slowly during the early stage of inflation. But the baryon density will be  diluted very fast. So the baryon density should be decided by the the chemical potential. At the late stage of inflation, the chemical potential will decrease strongly. The baryon density will be diluted much slower. The chemical potential can not produce baryon number any more. The baryon number will be frozen and the CP-violating interaction is decouple. It gives a natural interpretation for the decouple temperature of the CP-violating interaction Eq.~(\ref{rb}).

At the early stage of inflation, we can rewrite the Ricci scalar curvature and its time derivative as
\begin{align}
\mathcal{R}=6(2-\epsilon_1)H^2,\\
\dot{\mathcal{R}}=6\epsilon_1(2\epsilon_1-\epsilon_2-4)H^3,
\end{align}
where the $\epsilon_i$ are the Hubble flow-functions (HFFs)
\begin{align} \label{hff}
&\epsilon _1=-\frac{ d\ln H }{ d\ln a },
&\epsilon _{i+1}=\frac{ d\ln \epsilon _i}{d\ln a}.
\end{align}
In the vacuum inflation model, the quantum potential derive the inflation. Only $\epsilon _1$ is no trivial in this thoery. We can obtain
\begin{equation} \label{rsc}
	\mathcal{R}=-6pH^2,\dot{\mathcal{R}}=12(p+2)pH^3,
\end{equation}
where $p$ is the same parameter of Eq.~(\ref{hff}). From observation of cosmological perturbations we have $p+2=0.013$~\cite{0264-9381-34-10-105013}, which imply that the change rate of Ricci scalar curvature is not trivial during the inflation. At the late stage of inflation, the time derivative of the Ricci scalar curvature does not have a analytic form also. We need use numerical calculation. The relation between baryon density and chemical potential $\mu_B$ is~\cite{PhysRevD.74.087504}
\begin{equation}
	n_B=g_b\mu_B T^2/6,
\end{equation}
where $g_b\sim \mathcal{O}(1)$ is the number of intrinsic degrees of freedom of baryons.
In order to show the evolution of the baryon number, we need introduce the conformal baryon density $\eta _B\equiv n_B\times a^3$. The conformal baryon density reflect the baryon number indeed. As we showed in Fig.~(\ref{td}), the baryon number has a maximum value. This means the chemical potential can not produce baryon number after that scale. The baryon number will be stay at its maximum value other than decrease along the chemical potential. It is easy to see the decouple temperature will not influenced by parameter $M_*$. And the conformal baryon density is inversely proportional to $M_*^2$.
\begin{figure}[ht]
\centering
\includegraphics[width=0.45\textwidth]{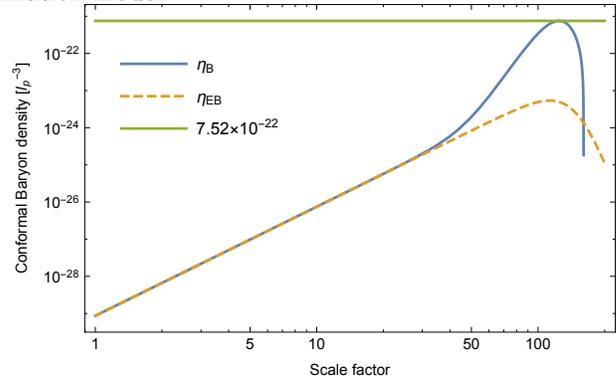}
\caption{The value of the conformal baryon density $\eta_{E}$ is provided by chemical potential. And $\eta_{EB}$ use asymptotic form Eq.~(\ref{rsc}). The figure uses the parameter $M_*=\sqrt{\frac{1}{8\pi}}$. We can see the conformal baryon density has a maximum expectation value.} \label{td}
\end{figure}

Once we obtain the decouple scale, we can calculate the ratio of baryons to photons at that temperature. For the entropy of the Universe is still increase until the end of the inflation, this asymmetry will be diluted. From numerical calculation, The dilution factor is approximately  $0.77$ and hence, the final asymmetry is
\begin{equation}
	\frac{n_B}{s}\approx 0.77 \frac{15g_b}{4\pi^2g_*}\frac{\dot{\mathcal{R}}}{M_* ^2 T}\bigg|_{T_D}.
\end{equation}

Now, only the cutoff scale $M_*$ remains uncertain. If we take it as reduced Planck scale  $M_*=\sqrt{\frac{1}{8\pi}}$, we will obtain $n_B/s\approx1.88\times10^{-10}$. It is a litter big than experimental result. This means the CP-violating interaction Eq.~(\ref{rb}) is able to give acceptable baryon asymmetry in the vacuum inflation model.

\section{DISCUSSION AND CONCLUSION}
We introduced the gravitational baryogenesis interaction in the vacuum inflation. The inflation derived by quantum potential will arise no trivial Ricci scalar curvature. Hence it will give a chemical potential for baryon. At the late stage of inflation, the quantum potential will decreases rapidly. The CP-violating interaction can no longer produce baryon number. The baryon number will be frozen. According to the calculation. This mechanism is able to  give acceptable baryon asymmetry.

\begin{acknowledgments}
This work is supported by the National Natural Science
Foundation of China (Grant Nos. 61471356).
\end{acknowledgments}

\end{document}